\documentclass[11pt]{article}
\usepackage{moriond,epsfig}
\usepackage{amssymb,latexsym}
\usepackage{amsmath}

\def\Journal#1#2#3#4{{#1} {\bf #2}, #3 (#4)}

\def\PLB{{\em Phys. Lett.}  B}
\def\PRL{\em Phys. Rev. Lett.}
\def\PRD{{\em Phys. Rev.} D}

\def\be{\begin{equation}}
\def\ee{\end{equation}}
\def\bea{\begin{eqnarray}}
\def\eea{\end{eqnarray}}

\begin{document}
\vspace*{4cm}
\title{Genericity of Cosmic Strings Formation in SUSY GUTs}

\author{J. ROCHER$^1$, R. JEANNEROT$^2$ and M. SAKELLARIADOU$^3$}

\address{$^1$Institut d'Astrophysique de Paris, {${\cal G}\mathbb{R}\varepsilon\mathbb{C}{\cal O}$}, 
FRE 2435, 98bis boulevard Arago, 75014 Paris, France\\
$^2$The Abdus Salam ICTP, strada costiera 11, 34100 Trieste, Italy\\
$^3$Division of Astrophysics, Astronomy, and Mechanics, Department of Physics,\\ 
University of Athens, Panepistimiopolis, GR-15784 Zografos (Athens), Hellas\\
and Institut d'Astrophysique de Paris, 98bis boulevard Arago, 75014 Paris, France}

\maketitle
\vspace{1cm}

\abstracts{
The idea of GUT implies that the universe went through a series of phase transitions during which 
topological defects are expected to form. We investigate here the genericity 
of cosmic strings formation in realistic SUSY GUTs models. We conclude that all acceptable 
symmetry breaking schemes produce cosmic strings after the last inflationary phase. 
Generically, as they form at the end of inflation, they have a mass of order of the GUT 
scale. Since cosmological data coming from CMB measurements do not show evidence for such 
strings, they constrain GUT scale physics.
}
\vspace{.5cm}

\section{Theoretical framework}
To describe the early universe, one has to take into account both particle physics and cosmology. 
High energy physics can provide an appropriate framework to study the early stages of our universe. 
Experiments in the two fields also lead to unavoidable constraints on theoretical models. Despite 
the fact that the Standard Model (SM), based on the gauge group $G_{\rm SM}\equiv SU(3)_{\rm 
C}\times SU(2)_{\rm L}\times U(1)_{\rm Y}$, has been tested to a very high precision at 
energies up to $\sim 200$ GeV, we know from neutrino oscillations~\cite{SK} that one has to go beyond 
it in order to describe physical processes above the electro-weak scale. 
Supersymetric (SUSY) Grand Unified Theories (GUTs) offer an appropriate framework for this goal.
They predict a unification of the three gauge couplings of the SM at a single point around $M_{\rm GUT}
\simeq 3\times 10^{16}$ GeV, by embedding $G_{\rm SM}$ into a unified gauge group $G_{\rm GUT}$ at high energies. 
They however imply proton decay~\cite{langacker} with a rate which can be made compatible with experimental 
limits: SUSY permits a unification at sufficiently high energies to be in agreement with the data. Such models 
contain many scalar fields which are natural candidates for the cosmologically relevant fields such as the 
inflaton. In addition, SUSY can also provide a natural candidate for dark matter, namely the Lightest 
SuperParticle (LSP).

In building SUSY GUTs models, one faces the problem of the appearance of undesirable topological 
defects such as monopoles or domain walls. Topological defects appear via the Kibble 
mechanism~\cite{kibble} during Spontaneous Symmetry Breaking (SSB). Their nature depends~\cite{ShelVil}  
on the topology of the vacuum manifold ${\cal M}=G/H$ when a gauge group $G$ spontaneously 
breaks down to a subgroup $H$. Due to the factor $U(1)_{\rm Y}$ of the SM, topologically stable 
monopoles always form between $G_{\rm GUT}$ and $G_{\rm SM}$ and would have overclosed the universe. 
This is the so-called ``monopole problem''. 

To solve the monopole, the flatness and the horizon problems one usually invokes a phase of inflation 
that occurs after the formation of the unwanted defects and thus dilutes 
them enough to make them compatible with cosmology. Note that this phase of inflation can also explain 
the fluctuations in the CMB. In SUSY GUTs, the most {\sl natural} model 
of inflation is hybrid inflation; the inflaton field is a gauge singlet which couples to a pair of GUT 
Higgs fields in complex conjugate representations which lowers the rank of the group by at least one unit. 
By {\sl natural}, 
we  mean that no extra-field, nor any extra-symmetry, are needed for inflation except those needed to 
build the GUT itself. This inflationary scenario can be of the F-term or D-term type depending on the 
origin of the dominant term during inflation. D-term inflation would necessitate the 
introduction of an additional $U(1)$ factor to $G_{\rm GUT}$ and we restrict our study to simple GUT groups.
Both standard F-term~\cite{Dvasha} and D-term inflation share the feature that the SSB induced by the Higgs 
fields ends the inflationary phase and thus no unwanted defects should form during this phase transition or 
later.

A successfull model must also be able to explain the matter/anti-matter asymmetry of the universe. 
This is naturally achieved in SUSY GUTs which predict massive neutrinos via the See-saw
mechanism. See-saw naturally gives rise to very light massive left handed neutrinos observed by 
experiments by introducing an additional SM gauge singlet (included in the GUT multiplet) which acquires 
a super-heavy mass through a Majorana mass term in the superpotential. 
The out-of-equilibium decay of heavy right-handed (s)neutrinos into electroweak Higgs(ino) and (s)lepton 
produces a net lepton asymmetry which is converted into a baryon asymmetry via sphaleron transitions. We 
make the asumption that baryogenesis occurs via leptogenesis. Therefore gauged $U(1)_{B-L} \subset G_{\rm GUT}$ 
should be broken at the end or after inflation.

Another ingredient coming from particle physics is the proton lifetime measurements~\cite{SK}, which 
suggest that a discrete $Z_2$ symmetry, called ``R-parity'' is unbroken at low energies. This symmetry 
protects the proton lifetime and also stabilizes the most interesting dark matter candidate: the LSP. 
The $Z_2$ symmetry contained in $U(1)_{B-L}$ can play the role of R-parity. We thus consider hereafter 
SSB patterns down to $G_{\rm SM}$ and $G_{\rm SM}\times Z_2$.

We considered in Ref.~\cite{jrs} a large number of simple Lie groups as the basis of our study since 
the unification of the three gauge couplings of the SM is automatic. To contain $G_{\rm SM}$, its rank 
has to be greater or equal to 4 and one must study the possible embeddings of $G_{\rm SM}$ in $G_{\rm GUT}$ 
to be in agreement with the SM phenomenology and especially the 
hypercharges ($Y$) of the known particles. We considered as many embeddings as we found in the 
litterature. The SM phenomenology imposes also that the group contains anomaly free complex representations 
for fermions. On the other hand, we restrict our study to groups of 
rank smaller than 8. We believe to capture the main results studying $SU(5)$, $SO(10)$, $SU(6)$, $SU(7)$, $E_6$, 
$SU(8)$, $SU(9)$ and $SO(14)$ and all possible SSB patterns down to $G_{\rm SM}$ and $G_{\rm SM}\times Z_2$. This 
list includes for example {\sl flipped} $SU(5)$ and $[SU(3)]^3$, as subgroups of $SO(10)$ and $E_6$ respectively.

\section{Spontaneous symmetry breaking patterns}
We list below the SSB schemes of some GUT groups down to $G_{\rm SM}$ and $G_{\rm SM}\times Z_2$, taken from 
Ref.~\cite{jrs} . We considered only regular subgroups that can be found in Ref.~\cite{slansky} and in~\cite{jrs} 
and only two discrete symmetries [the R-parity and the D-parity of charge conjugation contained in $SO(10)$] since 
the others have to be broken before inflation to avoid a domain wall problem. We shall denote by $\overset{n}
{\longrightarrow}$ a SSB during which there is formation of topological defects.  Their nature is given by $n$: 
$1$ for monopoles, $2$ for topological cosmic strings, $2'$ for embedded strings, $3$ for domain walls. We 
mention the formation of embedded strings~\cite{semiloc} despite the fact that they 
are not topologically stable and in general not dynamically stable either since mechanisms have been 
proposed to stabilize them. Moreover, despite their instabillity, they could play an interesting role 
in cosmology. Note also that $G \overset{1 ~(2)}{\longrightarrow} H~(Z_2)$ 
means $G \overset{1}{\longrightarrow} H$ and $G \overset{2}{\longrightarrow} H \times Z_2$. We refer 
the reader to Ref.~\cite{jrs} for more details.

\subsection{$SU(5)$} 
This group does not contain $B-L$ and it is unclear whether it is compatible with the proton lifetime 
measurements.
Since its rank is 4, it gives only one scheme $SU(5)\overset{1}{\longrightarrow} G_{\rm SM}$ where 
it is not possible to solve the monopole problem. Thus $SU(5)$ seems to be incompatible with cosmology provided 
our assumptions. 

\subsection{$SO(10)$} 
We list below the SSB schemes of $SO(10)$ via the Pati-Salam (PS) subgroup $G_{\rm PS}\equiv 
SU(4)_{\rm C} \times SU(2)_{\rm L}\times SU(2)_{\rm R}$. The two possible embeddings 
of the SM hypercharge ($Y$) in this subgroup are given by 
\begin{equation}
\frac{Y}{2} = \pm I_{\rm R}^3 + \frac{1}{2} (B-L)~,
\end{equation}
where $I_{\rm R}^3$ is the third generator of $SU(2)_{\rm R}$ and $B-L$ the generator of $U(1)_{B-L}$. 
Therefore, the breaking of the PS group down to $G_{\rm SM}~(\times Z_2)$ will form embedded strings.
Using the properties of the $SO(10)$ group~\cite{kibble2}, one can obtain the following SSB patterns :
\begin{equation}
\label{eq:ps}
\begin{array}{clllcccc} 
SO(10) \overset{1}{\longrightarrow}  &  4_{\rm C}     ~2_{\rm L}     ~2_{\rm R}   &  
\left\{ 
\begin{array}{cllllccc} 
\overset{1}{\longrightarrow} & 3_{\rm C} ~2_{\rm L} ~2_{\rm R} ~1_{\rm
B-L} & \left\{
\begin{array}{cllllccc}
 \overset{1}{\longrightarrow}  &   3_{\rm C}     ~2_{\rm L}     ~1_{\rm R}     ~1_{\rm B-L}   &  \overset{2 ~(2)}{\longrightarrow}  &   G_{\rm SM}  ~(  Z_2  ) \\ 
  \overset{2' ~(2)}{\longrightarrow}  &   G_{\rm SM} ~(Z_2)\\
 \end{array}
\right.
\\
  \overset{1}{\longrightarrow}  &   4_{\rm C}     ~2_{\rm L}     ~1_{\rm R}   &
\left\{ 
\begin{array}{cllllccc} 
  \overset{1}{\longrightarrow}  &   3_{\rm C}     ~2_{\rm L}     ~1_{\rm R}     ~1_{\rm B-L}   &   \overset{2 ~(2)}{\longrightarrow}  &   G_{\rm SM} ~(Z_2)\\
 \overset{2' ~(2)}{\longrightarrow}  &   G_{\rm SM} ~(Z_2)\\
 \end{array}
\right.
\\
  \overset{1}{\longrightarrow}  &   3_{\rm C}     ~2_{\rm L}     ~1_{\rm R}     ~1_{\rm B-L}   &  ~~~~\overset{2 ~(2)}{\longrightarrow}      ~~G_{\rm SM} ~(Z_2)\\
  \overset{1 ~(1,2)}{\longrightarrow}  &   G_{\rm SM}  ~(  Z_2  )\\
  \end{array}
\right.
\end{array}
\end{equation}

From the above schemes, one can see that the first five patterns can give rise to models that solve 
the monopole problem since the inflationary phase can take place at the beginning of the last SSB. Note 
that there is no other choice since inflation has to take place after the last formation of monopoles. 
In addition, the SSB that ends inflation always produces a network of strings of the GUT scale, sometimes 
embedded and sometimes topologically stable depending on the group unbroken at low energy. If we impose 
that the scheme ends with $G_{\rm SM}\times Z_2$, then the strings are always topological. The sixth 
pattern is incompatible with cosmological data since the only SSB forms monopoles and is therefore ruled out.

Note that, to obtain all possible SSB patterns from $SO(10)$ down to  $G_{\rm SM} (\times Z_2)$, 
one has to consider schemes via the subgroup $SU(5)$, those through $G_{\rm PS}\times Z_2^C$ 
where $Z_2^C$ is the D-parity and the more direct ones. They are exhaustively listed in Ref.~\cite{jrs}. 
To conclude with the $SO(10)$ group, we obtain 68 schemes where the monopole problem can be solved. 
Among them only 34 can satisfy all the constraints and they all give rise to topological strings 
formation at the end of the inflation phase, sometimes accompagnied by embedded strings.

\subsection{Higher order groups}
Considering other GUT groups, the minimal $SU(6)$ and $SU(7)$ do not contain as a subgroup $U(1)_{B-L}$. 
They are thus incompatible with high energy physics phenomenology. $E_6$ gives rise to 1268 schemes where 
it is possible to solve the monopole problem. All of them are listed in Ref.~\cite{jrs}. Among them, 676 
let the R-parity unbroken at low energy and 534 satisfy all constraints, and all form cosmic strings after 
inflation. 
The reason is that successfull SSB patterns look often similar to those of the previous example: the 
monopole are formed in the begining, before inflation, and then the last SSBs are of the form:  $3_{\rm C}
~2_{\rm L}~1_{\rm R}~1_{\rm B-L}\overset{2 ~(2)}{\longrightarrow}G_{\rm SM} ~(Z_2)$.
For $SU(8)$, $SU(9)$ and $SO(14)$, the conclusion remain unchanged if one imposes that the $B-L$ symmetry 
has to be broken after inflation. Note that if we relax this constraint, the conclusions remain almost 
unchanged: approximately all the SSB patterns give rise to cosmic strings formation after the inflation phase.

\section{Conclusion}
We considered here the formation of topological defects during the SSB patterns of GUT gauge groups. 
We selected SUSY models that take into account cosmological and particle physics constraints. 
We found that the formation of cosmic strings after inflation is very generic: within our assumptions, 
they are unavoidable. They generically form at the end of inflation and have therefore a mass per unit 
length proportional to the inflationary scale (especially for small groups of rank 5 or 6). 
They would therefore be expected to have an important cosmological role, in apparent contradiction with CMB 
data. This leads in fact to strong constraints on particle physics model parameters~\cite{prl}.

\section*{Acknowledgments}
It is a pleasure to thank P. Peter for usefull comments and the organisers of the Moriond meeting for inviting me 
to give this talk.

\end{document}